\documentclass{article}
\usepackage{amssymb,amsfonts,amsmath,amsthm}
\usepackage{graphicx} % Required for inserting images

\usepackage[maxbibnames=99,url=false]{biblatex}
\usepackage{bm}
\usepackage{amssymb}
\usepackage{amsmath}
\usepackage[hidelinks]{hyperref}
\usepackage{booktabs}
\usepackage{units}
\usepackage{braket}
\usepackage{algorithm}
\usepackage{algpseudocode}

\newcommand{\T}{\intercal}

\newcommand{\R}{\mathbb{R}}

\newcommand{\C}{\mathbb{C}}
\newcommand{\I}{\mathbb{I}}

\theoremstyle{definition}
\newtheorem{definition}{Definition}

\title{Quantum algorithms for \textit{N-1} security in power grids}
\author{Niels M. P. Neumann\footnote{The Netherlands Organisation for Applied Scientific Research (TNO), 2595DA The Hague, The Netherlands} \and Stan van der Linde$^*$ \and Willem de Kok$^*$ \and Koen Leijnse\footnote{Quantum Application Lab (QAL), Amsterdam, The Netherlands} \and Juan Boschero$^*$ \and Esteban Aguilera$^*$ \and Peter Elias-van den Berg$^*$ \and Vincent Koppen\footnote{Alliander Research Center for Digital Technologies, P.O. 50, 6920 AB Duiven, The Netherlands} \and Nikki Jaspers$^{\ddag}$ \and Jelte Zwetsloot$^{\ddag}$}

\begin{document}

\maketitle

\begin{abstract}
%With the energy transition ahead of us and with new ways of generating energy becoming more widespread, energy grid operators face new challenges. 
%New infrastructure is needed, while ensuring that both existing and new infrastructure can withstand unforeseen failures. 
%Energy grid operators can use these quantum approaches to solve the computationally hard \textit{N-1}-security question.
In recent years, the supply and demand of electricity has significantly increased.
As a result, the interconnecting grid infrastructure has required (and will
continue to require) further expansion, while allowing for rapid resolution
of unforeseen failures.
Energy grid operators strive for networks that satisfy different levels of security requirements.
In the case of \textit{N-1} security for medium voltage networks, the goal is to ensure the continued provision of electricity in the event of a single-link failure.
However, the process of determining if networks are \textit{N-1} secure is known to scale polynomially
in the network size. 
This poses restrictions if we increase our demand of the network. 
In that case, more computationally hard cases will occur in practice and the computation time also increases significantly. 
In this work, we explore the potential of quantum computers to provide a more scalable solution.
In particular, we consider gate-based quantum computing, quantum annealing, and photonic quantum computing.
\end{abstract}

\section{Introduction}
%Multiple challenges lie ahead of us in the coming years, with the energy transition being one of them. 
%With the energy transition also come new ways of generating electricity, such as residential solar panels. 
%Households therefore act as both an energy provider and consumer, depending on, for instance, the weather conditions. 
%A suitable and robust infrastructure proves vital to accommodate for these different usages. 

In the past years, the rapid economic growth, energy transition, and digitization of our society and industry
has resulted in a steady increase of the electricity consumption.
In particular, there has been a significant rise in the energy demand originating from new offices, factories,
residential areas, data centers, electric vehicles, and electric heat pumps.
In order to meet this demand, the supply has, in turn, increased (partly in a decentralized manner),
which has been made possible by the utilization of solar/wind farms, local energy communities, as well as energy imports.

Naturally, supply and demand can be matched insofar as the interconnecting infrastructure is efficiently and safely operated, troubleshooted, reinforced, and expanded.
%This requires the networks to be sufficiently large to provide all end-users with the requested energy, while also being robust against possible failures. 
For Medium Voltage (MV) networks, these constraints translate to the so-called \textit{N-1} security, a security constraint that ensures continuous supply even in cases where a single link in the network fails. 
Checking the \textit{N-1} security of MV networks translates into addressing two questions:
First, does the given network allow for a valid reconfiguration of the network in the event of any single-link failure?
Second, how can we optimally reconfigure our network such that the network constraints remain satisfied?

Currently, there exist algorithms to efficiently check if a reconfiguration of a network is valid and can even determine how to reconfigure a network optimally in simple cases. 
In harder cases, current algorithms already take longer. 
Checking if the network as a whole is \textit{N-1} secure requires performing these computations for every possible single edge failure. 
As the demands on MV networks increase due to higher usage, more hard reconfigurations will be found and the running time of the algorithm will increase. 
Even small improvements in the computations for the hard cases will directly affect the security of networks and help prepare the electricty grid for the future. 
%For a small group of networks, however, these algorithms tend to take longer before providing an answer. 
%For a single instance, this is doable;
%however, when having to run the algorithm often, even small improvements in the algorithm running times can give significant advantages.

In the last decades, research on quantum computers has led to identifying
theoretical computational advantages in terms of time complexity~\cite{Grover:1996,Shor:1997,Lanyon2010,HHL:2009}. 
%Quantum computers use the laws of quantum mechanics to perform computations and solve problems. 
%These novel devices can, at least in theory, offer significant improvements in specific fields
Nonetheless, improvements in theory do not necessarily translate to improvements in practice, as is the case for the HHL algorithm as discussed in~\cite{Aaronson2015}.
Furthermore, quantum computers are still at a stage of active development and a definitive way in which to perform quantum computations remains unclear. 
However, numerous organizations worldwide have already started to prepare for the advent of scalable quantum computers.
To this end, critical processes, such as those related to \textit{N-1} security, are currently being reviewed to provide a timely assessment of the potential benefits of quantum technology.
%Yet, quantum computers offer new ways to perform computations which can, especially for computationally heavy problems, have great impact. 

%In this work, we explore how quantum computers can help solving the challenges encountered in measuring network robustness. 
In this work, we explore multiple approaches to implementing quantum algorithms for
\textit{N-1} security.
In Section~\ref{sec:preliminaries}, we provide a short introduction to quantum computing,
formally define the \textit{N-1} problem, and discuss a classical algorithm.  
In Sections~\ref{sec:gate_based} and~\ref{sec:qannealing},
we describe a gate-based quantum approach and a quantum annealing approach.
Next, the potential of photonic quantum computing is discussed in Section~\ref{sec:photonics}.
Conclusions are presented in Section~\ref{sec:conclusion}.
%We present a gate-based approach based on Grover's search algorithm~\cite{Grover:1996},
%as well as a QUBO formulation and implemention on quantum annealing hardware.
%In addition, we discuss the potential of photonic quantum computing.

%Each of these three approaches tries to tackle the \textit{N-1} security problem, which we formally discuss in Section~\ref{sec:def_n1}. 
%For each, we briefly introduce that type of quantum computing, the potential of the quantum approach and present the algorithm and discuss the results (only for gate-based and annealing).

\section{Preliminaries}
\label{sec:preliminaries}
\subsection{Quantum computing}
From an abstract viewpoint, conventional computers work by manipulating bits,
each of which represents a logical state as indicated with either~$0$ or~$1$.
By applying operations to those bits, computers are able to run programs
that allow us to solve problems.
Typically, electric currents are used to implement bits while transistors are used to implement operations. 
Quantum computers work in a similar fashion by making use of quantum mechanical principles.
Usually, two-state quantum mechanical systems are used, called qubits, by analogy with bits. In addition, there are two properties that give quantum computers their power: \textit{superposition} and \textit{entanglement}, which are described below. 

Superposition is the property that a quantum state can be indefinite,
as opposed to bits in conventional computers, which always have a single definite state. 
For the definite states~$\ket{0}$ and~$\ket{1}$, which are the quantum equivalent
of the bits~$0$ and~$1$,
a quantum state in superposition is given by $\ket{\psi}=\alpha\ket{0} + \beta\ket{1}$,
where $\alpha$ and $\beta$ are complex numbers and $|\alpha|^2+|\beta|^2=1$. 
Operations can be performed on the superposition and only upon measurement of the state will one of the two definite states be found. 
The probability of finding the system in a specific definite state
is equal to the associated amplitude squared. 

Entanglement is the property that two quantum states can be correlated beyond what is classically possible.
A famous example of an entangled state is the GHZ state~\cite{greenberger1989bell}: 
\begin{equation*}
    \frac{1}{\sqrt{2}}(\ket{0}_A\ket{0}_B + \ket{1}_A\ket{1}_B).
\end{equation*}
Two parties, $A$ and $B$, each have a single qubit. 
When viewed locally, each qubit is in a uniform superposition. 
Globally, however, the two qubits are entangled such that the state of one always matches the state of the other qubit. 
Hence, a measurement on either of the states gives one of the two states with equal probability. 
However, measuring one party's qubit directly determines the state of the other party's qubit. 

The task of quantum algorithm developers is to leverage these two effects in such
a way that desirable measurement outcomes are promoted,
while undesirable measurement outcomes are suppressed.
%Quantum hardware is still under rapid development and the available quantum resources differ both in number and in quality for different hardware technologies and quantum computing approaches. 

\subsection{The \textit{N-1} problem}\label{sec:def_n1}
\begin{figure}
    \centering
    \includegraphics[width=0.33\textwidth]{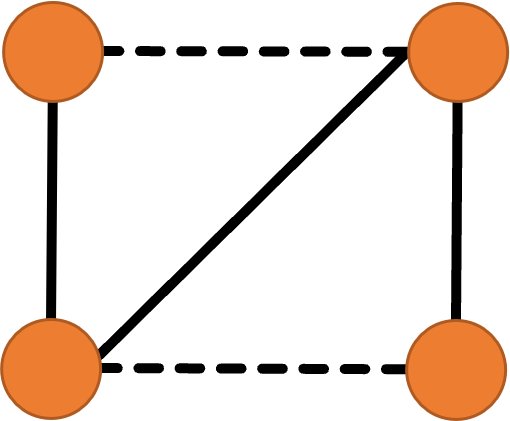}
    \qquad
    \includegraphics[width=0.33\textwidth]{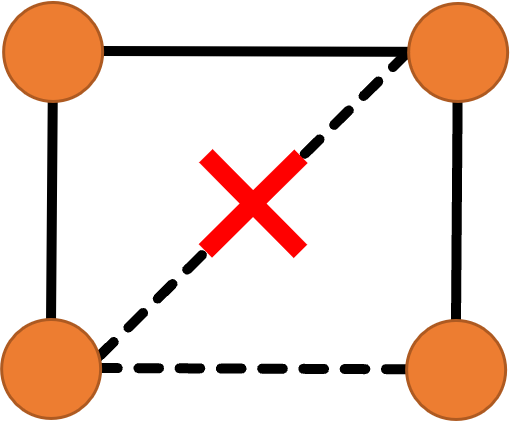}
    \caption{Left: An energy grid with solid edges representing active cables and dashed edges representing inactive cables. 
    Right: Upon edge failure, we turn on an inactive edge to again obtain a spanning tree.}
    \label{fig:simple_network_and_failure}
\end{figure}

Electrical grids are naturally modeled using graphs,
where (secondary) substations are represented by nodes, and cables are represented by edges.
Cables are assumed to be in one of two states, either \textit{active} or \textit{inactive}.
Active cables are used to transport energy, while inactive cables
are there for redundancy purposes and can be activated in the event of a failure of a currently
active cable. In this work, we assume that the graph with active edges always forms a spanning tree,
which we refer to as a \textit{configuration}.
\autoref{fig:simple_network_and_failure}(Left) shows a simple network with
three active edges (solid lines) and two inactive edges (dashed lines). 
When one edge fails, as indicated by the cross in \autoref{fig:simple_network_and_failure}(Right),
a possible reconfiguration is found by deactivating the failing edge, and activating
an inactive edge.
We refer to these events individually as a \textit{switch}.
When considered together, we use the term \textit{switchover}.

From a graph-theoretic viewpoint, finding a new spanning tree directly leads to a reconfiguration.
However, since we are dealing with electricity networks, additional constraints must be imposed
for the reconfiguration to be valid.
In particular, nodes (substations) are required to comply with specific voltage limits,
while edges (cables) must comply with current limits.
These constraints are jointly referred to as \emph{load-flow} constraints.
A direct consequence of adding these constraints is that a single switchover
is oftentimes insufficient to meet all constraints. 
As the complexity of the algorithm increase with the number of switchovers, in practice a limit $k$ is usually imposed on it. 
The goal therefore is to find a valid reconfiguration using at most $k$ switchovers.
%The goal is then to find the minimum number of switchovers that yield a valid reconfiguration.

We call a graph \textit{N-1} secure if a valid reconfiguration exists for every possible single edge failure. 
\begin{definition}[\textit{N-1} security]
    Consider a graph $G=(V, E_{a}\cup E_{i})$, consisting of active $E_a$ and inactive edges $E_i$.
    We call the graph $G$ \textit{N-1} secure with switchover parameter $k$ if for any single edge failure, a valid reconfiguration exists with at most $k$ switchovers, such that the reconfiguration meets the load-flow constraints.
\end{definition}
In practice, $k=1$ suffices for most active edges in a graph. 
These can be easily found, as described in the next section. 
However, the remaining edges require a larger $k$, which increases the solution space and running time
of the algorithm that searches for valid reconfigurations.
 
\subsection{Classical algorithm}
For the classical algorithm, we consider a two-step algorithm. 
Though the exact implementation might differ, we assume this algorithm for comparison with the quantum algorithms. %The classical algorithm has two parts.
First, we exhaustively list all potentially valid reconfigurations which are $k=1$ switchovers away from the current one.
Usually, that list will contain valid reconfigurations that apply to 90\% of the active edges, and we would therefore know which switches to apply in the event that any of those edges fails.
Second, we exhaustively list potentially valid reconfigurations that meet the following two conditions:
1) they deactivate at least one of the remaining 10\% of the active edges, and 
2) they are $k>1$ switchovers away from the current one.
When (if at all) a valid reconfiguration has been found for each remaining active edge (by possibly increasingly trying larger values of $k>1$ when necessary for some active edges), we can confirm that the network is \textit{N-1} secure and know exactly how it must be reconfigured in the event of a single failure.
This two-step algorithm is described in detail in Algorithm~\ref{alg:classical_step1} (step 1) and Algorithm~\ref{alg:classical_step2} (step 2), which, in turn, make use of Algorithm~\ref{alg:enumerating_trees} for enumerating spanning trees.
As mentioned, further classical optimizations are possible, yet, the focus is on the high-level idea behind the algorithm. 
Note that in Algorithm~\ref{alg:enumerating_trees}, in line~\ref{alg:enumerating_trees:active_edges} we can force the choice of $k$ active edges to guarantee a spanning tree.

\begin{algorithm}
    \caption{Classical algorithm: step 1}
 \begin{algorithmic}[1]
      \State $L \leftarrow$ Algorithm \ref{alg:enumerating_trees} with $k=1$ (without providing the parameter $D$)
      \Comment{Generate a list $L$ of all possible reconfigurations which are $k$ switchovers away from current configuration.}
      \State For each reconfiguration $l$ in $L$, determine which active edge $e$ is deactivated.
      \For{each $(e, l)$ pair found in the previous step}
        \State Perform load-flow check on $l$.
        \If{load-flow check passes}
            \State Save $(e, l)$.
            \Comment{This indicates that $l$ can be used as reconfiguration if $e$ fails.}
            \State Continue with the next edge in the for-loop.
        \EndIf
      \EndFor
      \Comment{At the end of this step, $\approx90$\% of the edges that are allowed to fail have been identified.}
 \end{algorithmic}
 \label{alg:classical_step1}
\end{algorithm}

\begin{algorithm}
    \caption{Classical algorithm: step 2}
 \begin{algorithmic}[1]
      \State $L \leftarrow$ Algorithm \ref{alg:enumerating_trees} with $k>2$ and $D$ set to the remaining 10\% of the edges.
      \Comment{Note that each spanning tree in $L$ can be a valid reconfiguration for more than one failing edge.}
     \For{each active edge $e$ in $D$}
         \If{we already have a pair $(e, l)$}
         \Comment{If a reconfiguration has been found in a previous iteration while searching for reconfigurations for a different active edge.}
             \State Continue with the next edge.
         \EndIf
         \For{each reconfiguration $l$ in $L$}
             \If{$e$ is deactivated in $l$}
                 \State Perform load-flow check on $l$.
                 \If{load-flow check passes}
                     \For{each edge $e_d$ deactivated in $l$}
                         \State Save $(e_d, l)$.
                         \Comment{This indicates that $l$ can be used as reconfiguration if $e_d$ fails.}
                     \EndFor
                     \State Continue with the next edge in the outer for-loop.
                 \EndIf
             \EndIf
    \EndFor
    \EndFor
 \end{algorithmic}
 \label{alg:classical_step2}
\end{algorithm}

\begin{algorithm}
    \caption{Algorithm to enumerate all spanning trees for an arbitrary~$k$}
    \textbf{Input}:
    \begin{enumerate}
        \item A graph (current configuration and inactive edges).
        \item $k$: number of switchovers.
        \item (Optional) A list of specific active edges $D$.
    \end{enumerate}
    \textbf{Output}:
    A list of spanning trees. Note: if $D$ has been specified, the output list will only include those spanning trees that deactivate at least one edge in $D$. 
 \begin{algorithmic}[1]
      \State Create an empty list ($S$) of spanning trees. 
      \State Based on the input graph, create a table $T$ where each row contains:
      \textbf{Column 1}: An inactive edge;
      \textbf{Column 2}: A fundamental cycle (represented by a set of active edges) corresponding to the inactive edge in \textbf{Column 1}.
      \For{each combination of $k$ rows from $T$}
      \State Use those rows to form a new table (of $k$ rows and $2$ columns) called $U$.
      \For{$1\leq i \leq k$}
          \State $x_i \leftarrow$ inactive edge in row $i$ and \textbf{Column 1} of $U$.
          \State $C_i \leftarrow$ set of actives edges in the fundamental cycle in row $i$ and \textbf{Column 2} of $U$.
      \EndFor
      \State Generate list of all possible combinations of $k$ active edges. Each combination is given by $(e_1,e_2,\cdots,e_{k}) \in C_1 \times C_2 \times \cdots \times C_{k}$.\label{alg:enumerating_trees:active_edges}
      \For{each combination of the form $(e_1,e_2,\cdots,e_{k})$}
      \If{$D$ has been provided and $D \cap (e_1,e_2,\cdots,e_{k}) = \emptyset$}
      \State Continue with the next combination $(e_1,e_2,\cdots,e_{k})$.
      \EndIf
      \State Generate a new configuration by activating $x_i$, (for all $i\leq k$) while deactivating $(e_1,\cdots,e_{k})$.
      \State Add the found configuration to $S$.
      \EndFor
      \EndFor
      \State \Return $S$.
 \end{algorithmic}
 \label{alg:enumerating_trees}
\end{algorithm}

\section{Gate-based quantum computing approach}
\label{sec:gate_based}
\begin{figure}[tbp]
\centering
\includegraphics[width=0.4\textwidth]{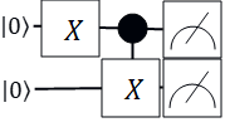}
\caption{An example quantum circuit. 
Horizontal lines indicate qubits. Time flows from left to right. 
The squares indicate quantum gates, and the black dot represents a controlled operation. 
The squares with the meter symbol indicate a measurement of the qubit.
This circuit returns the same measurement result for both qubits with certainty.}
\label{fig:qb_example_circuit}
\end{figure}

Gate-based quantum computers are most similar to conventional quantum computers: qubits are manipulated by quantum gates.
Quantum algorithms are often represented by quantum circuits. 
\autoref{fig:qb_example_circuit} gives an example quantum circuit, where every line represents a qubit and time flows from left to right. 
%Other models of universal quantum computing also exist, such as measurement-based quantum computing used in photonics, or adiabatic quantum computing, closely related to quantum annealing. Both are equivalent to gate-based quantum computing in their powers, with a potential constant overhead~\cite{aharonov2005adiabatic,knill2000efficient}. Most physical implementations of quantum computers are however based on the gate-based model, due to its relative ease for algorithm design. Circuit operations can be described mathematically using linear algebra and are often represented graphically using circuit diagrams, with the qubits in their starting state on the left and gates applied reading left to right. On the very right, measurements are performed to extract information, forcing the measured qubits to collapse to a computational basis state; a 0 or a 1. An example quantum circuit is shown in Figure \ref{fig:qb_example_circuit}. 

The problem of finding reconfigurations is similar to unstructured search:
The load-flow constraints make that, at first sight, it is hard to determine if reconfigurations are indeed valid. 
The famous Grover's quantum algorithm improves the complexity of unstructured search from linear ($\mathcal{O}(N)$) to squared root ($\mathcal{O}(\sqrt{N})$)~\cite{Grover:1996}. 
Giving a quadratic improvement over classical alternatives. 
The interesting fact here is that simply listing a database already costs linear time, yet, quantum computers can search through them quadratically faster. 

With this algorithm, iteratively, the amplitudes corresponding to valid reconfigurations are boosted and those of invalid reconfigurations are suppressed. 
The Amplitude Amplification subroutine plays a crucial role in Grover's algorithm. 

\subsection{Amplitude Amplification for the \textit{N-1} problem} 
The Amplitude Amplification algorithm consists of two alternating operations applied to an initial superposition that covers the entire dataset. 
The first operation marks the good states by multiplying the phase of the corresponding quantum states by minus one. 
The second operation inverses the amplitudes over the averages of the amplitudes of all states. 
As some states have a negative amplitude, the average amplitude is smaller than that of the bad states.
After one such alternation, the good states have a slightly larger amplitude than the bad states.
Repeating this procedure assures that the amplitude of the good states increases, whereas the bad states have an amplitude that approaches zero. 
 
The first operation is usually referred to as the oracle step. 
In theoretical works, this oracle is often assumed as a black box, whereas in practice, the oracle has to be explicitly implemented. 
Due to the limited quantum resources and the difficulty of implementing the load-flow constraints in a quantum circuit, we restrict ourselves to query complexity. 
 
In the second operation, the amplitudes are inverted over the mean of the amplitudes.
A key part of this operation is reverting the initial superposition and then again preparing it.
The initial superposition must be easily obtainable, as the construction is also implicitly used in this second operation.  

One iteration of the algorithm consists of the successive application of these two operations. 
The optimal number of iterations follows from the total size of the dataset and the number of good states, yet even if the number of good states is unknown, with high probability a marked state is found, with a constant overhead in the complexity~\cite{Boyer_1998}.  

To implement the algorithm for the \textit{N-1} problem, we need to prepare a useful initial superposition over possible reconfigurations. 
For that, we use an identification number that we link to different reconfigurations. 
We use the same identification number to implement the oracle that identifies the valid reconfigurations. 

\subsection{Implementation and results} 
We use the quantum algorithm specifically for the $k>1$ case, as $k=1$ is already solved efficiently classically. 
We made an implementation in Python using the Qiskit module by IBM~\cite{qiskit2023}. 
The implementation can be run on either a classical simulator or on actual quantum hardware and is available online~\cite{TNO_Quantum_N1_problem_2023}.  

As input, the algorithm takes a network, the failing edge and the number of switchovers. 
For the proof-of-principle, we also used the link between identification number and reconfiguration, as well as information on which reconfigurations are indeed valid. 

\begin{figure}[tbp]
\centering
\includegraphics[width=0.8\linewidth]{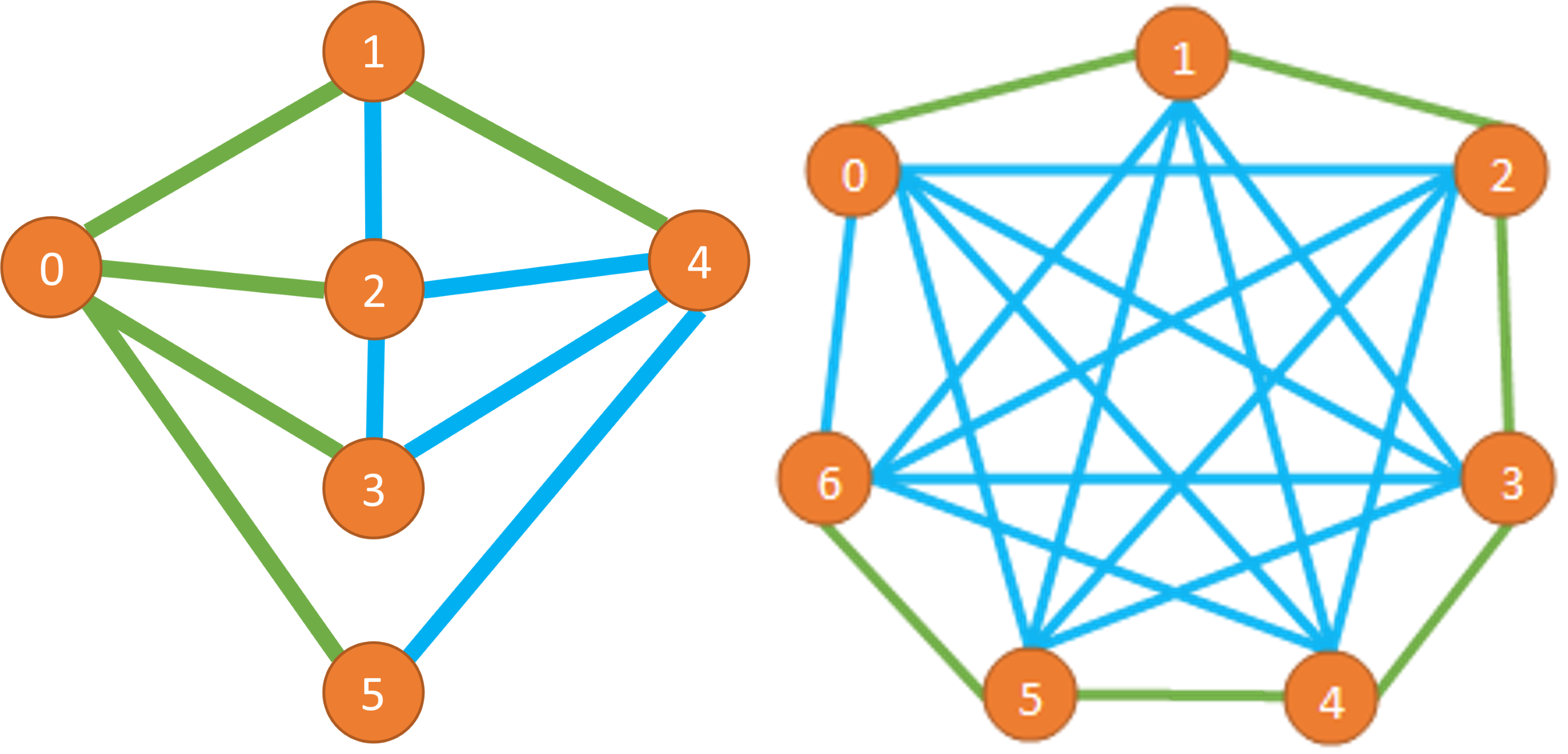}
\caption{The two graphs used in the experiments for the gate-based implementation.
Green edges represent active edges, whereas blue edges represent inactive edges.}
\label{fig:gb_networks}
\end{figure}

We have tested the algorithm using the two graphs shown in \autoref{fig:gb_networks}.
The green edges are active edges and the blue edges are inactive edges.  

\begin{figure}[tbp]
\centering
\includegraphics[width=0.53\linewidth]{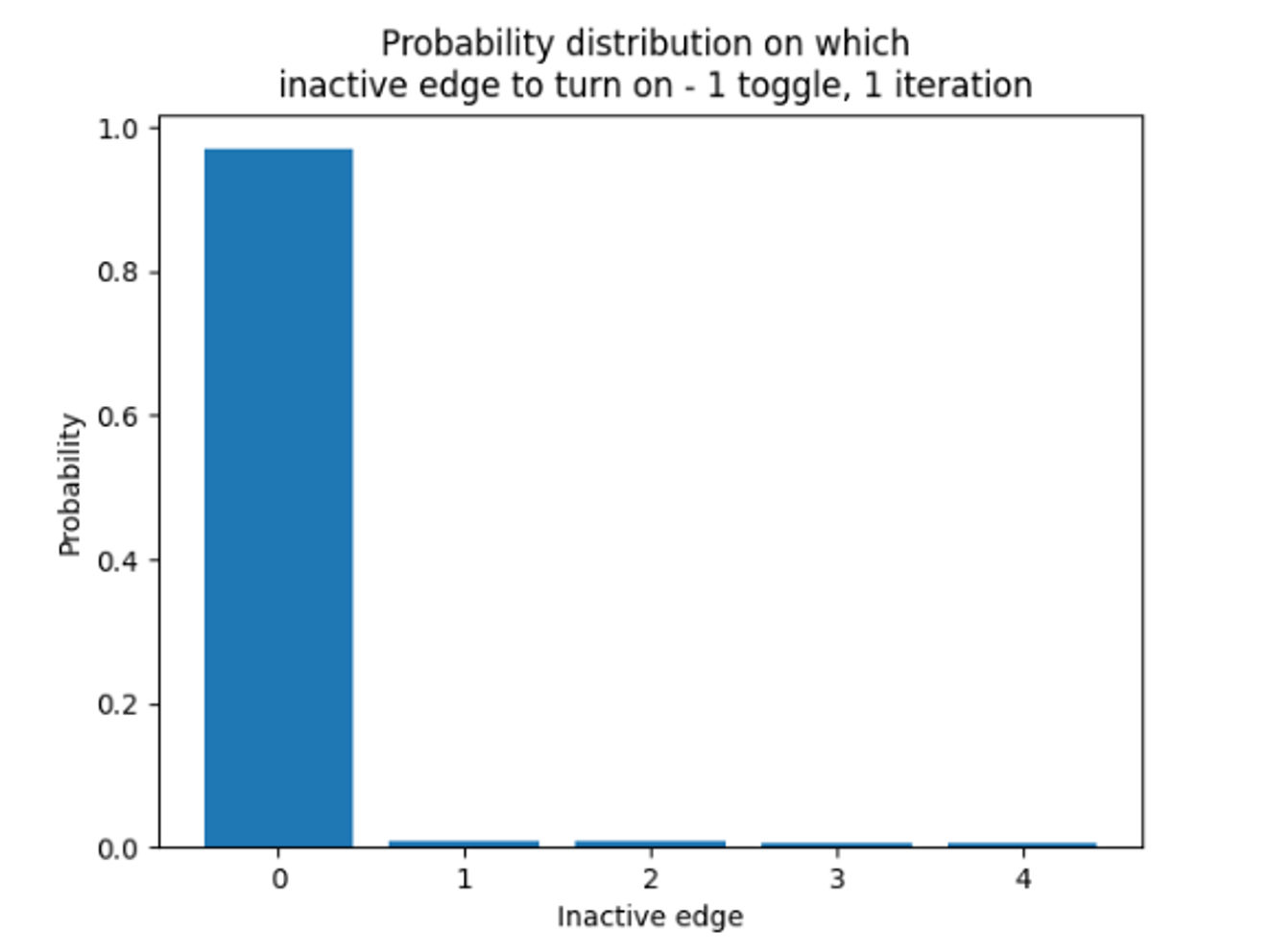}
\caption{Results of the first experiment with one failing edge and one valid reconfiguration.}
\label{fig:gb_exp_1}
\end{figure}

In the first experiment, we let edge $(0,2)$ fail and we consider only a single switchover.
The only valid reconfiguration would be to turn edge $(1,2)$ on and the output of the quantum algorithm is given in \autoref{fig:gb_exp_1}.
We see that with high probability the quantum algorithm returns the zeroth inactive edge to turn on, which corresponds precisely with the $(1,2)$ edge as desired.  

\begin{figure}[tbp]
\centering
\includegraphics[width=1\linewidth]{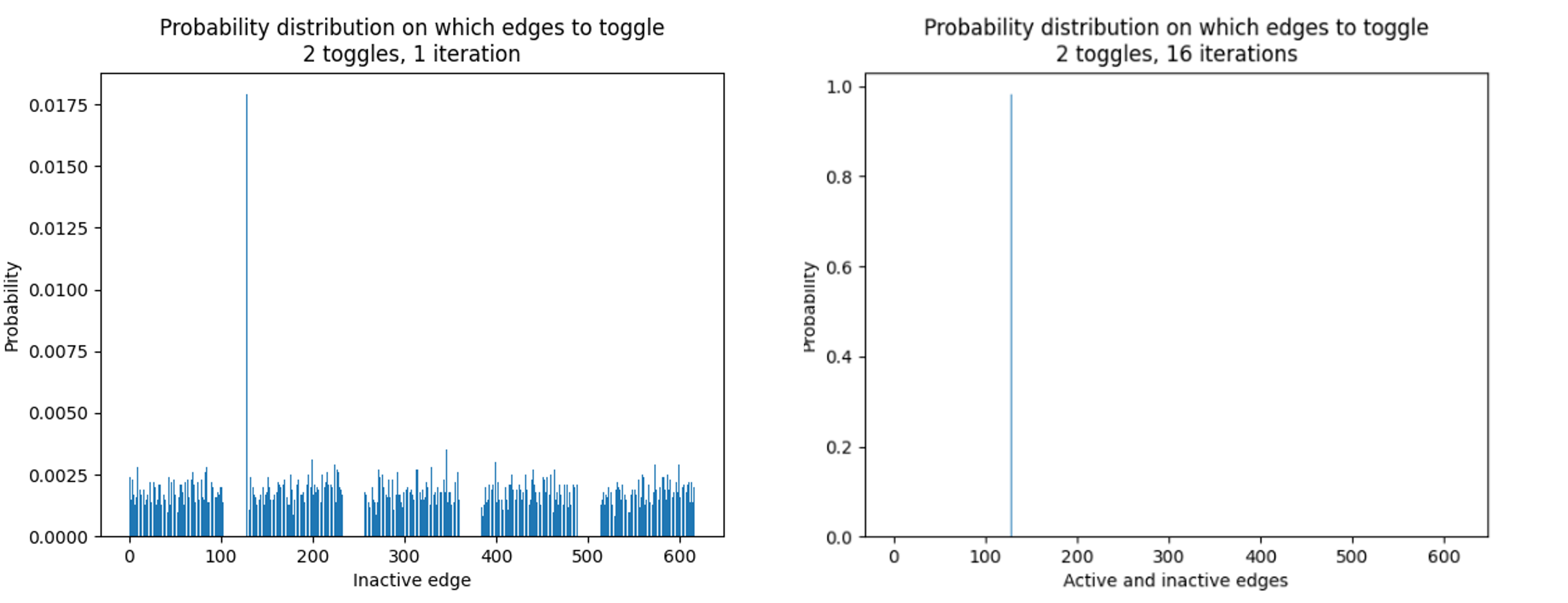}
\caption{Results of the second experiment with one failing edge and one valid reconfiguration after a single iteration and after sixteen iterations.}
\label{fig:gb_exp_2}
\end{figure}

In the second experiment, we consider the right graph of \autoref{fig:gb_networks}
We now let the $(1,2)$ edge fail and consider the case of two switches. 
In case of only a single valid solution with two switchovers, the valid network requires edges $(0,2)$ and $(0,3)$ to be turned on and the edge $(2,3)$ to be turned off.
The probability distribution after a single iteration approximately looks like the graph shown in \autoref{fig:gb_exp_2} (left). 
We see a single state with a higher probability, which corresponds precisely to the valid reconfiguration. 
The probability of finding this specific reconfiguration is however still low.
If we increase the number of iterations to 16, we see that the probability of correctly finding this reconfiguration approaches one, as shown in the same figure on the right. 

\begin{figure}[tbp]
\centering
\includegraphics[width=0.53\linewidth]{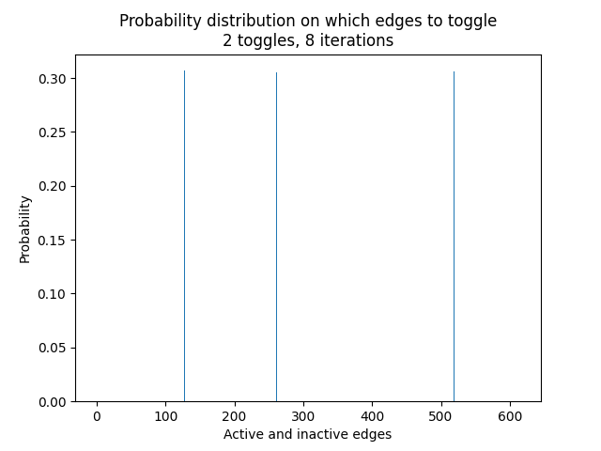}
\caption{Results of the third experiment with one failing edge and three valid reconfigurations after eight iterations.}
\label{fig:gb_exp_2_fe}
\end{figure}

In the previous experiment, we had only a single valid reconfiguration. 
We finally consider a situation with three valid reconfigurations. 
The one from the previous experiment, one with edges $(0,2)$ and $(1,4)$ on and edge $(3,4)$ off, and one with edges $(0,2)$ and $(1,6)$ on and edge $(5,6)$ off. 
As more valid reconfigurations exist, the number of iterations required to find a valid one decreases.
\autoref{fig:gb_exp_2_fe} shows the probability distribution after eight iterations with three valid reconfigurations all having approximately the same probability of being found.  

The used quantum gate-based approach requires quadratically fewer queries to the load-flow computations compared to what a classical unstructured search requires. 

\section{Quantum annealing approach}\label{sec:qannealing}
\subsection{Introduction to quantum annealing}
Within adiabatic quantum computing, a system is initialised in a simple ground state and then slowly modified to a different ground state, corresponding to the answer of the problem~\cite{farhi2000quantum}.
If this change is slow enough, the system remains in the ground state and measuring the final state answers the problem~\cite{griffiths2018introduction, messiah2014quantum}. 
If the change is too fast, the system can leave the ground state and different answers can be found. 
In practice, this is achieved by applying a simple Hamiltonian on the system to obtain the initial ground state. 
Next, this simple Hamiltonian is evolved adiabatically (slowly) to a Hamiltonian whose ground state describes the answer to the posed question.
Because the evolution of the Hamiltonian happens adiabatically, the adiabatic theorem guarantees that the state after the evolution is the ground state of the final Hamiltonian.

If the evolution happens too fast, the system can leave the ground state and different measurement outcomes are found. 
\citeauthor{kadowaki1998quantum} were the first to mention quantum annealing~\cite{kadowaki1998quantum}. 
They drew inspiration from the classical meta-heuristic simulated annealing~\cite{kirkpatrick1983optimization}, which in turn lends its name from the annealing process in metallurgy, in which materials are heated and cooled in a controlled manner to change its physical properties.
Simulated annealing proves useful for a wide range of optimisation problems, including Quadratic Unconstrained Binary Optimization (QUBO) problems.
It works by searching the solution space, while slowly decreasing the probability of accepting worse solutions than the current solution.
For a more in-depth analysis of the work, we refer to~\cite{henderson2003theory}.
	
Quantum annealing was not proposed as alternative to adiabatic quantum computing in mind.
Yet, we can interpret quantum annealing appears as a heuristic version of adiabatic quantum computing with two key differences~\cite{mcgeoch2022adiabatic}:
\begin{itemize}
    \item The final Hamiltonian in quantum annealing represents a classical discrete optimisation problem (usually a NP-Hard problem), versus any arbitrary problem in adiabatic quantum computing;
    \item Quantum annealing evolves the system faster and thereby allows the system to leave the ground state.
\end{itemize}
Since the problem we study in this use case can be formulated as a discrete optimisation problem, we can turn to quantum annealing to find solutions.
Note that quantum annealing gives no guarantees on the success probability, whereas adiabatic quantum computing does so, provided a slow enough evolution. 
In practice, the system has a non-zero (often significant) probability to leave the ground state.
This is mitigated by running a quantum annealing computation multiple times, which additionally also gives a heuristic version of a probability distribution.

\subsection{The D-Wave implementation and QUBOs}
Quantum annealing computers (or quantum annealers) can solve problems using quantum annealing.
Different companies offer quantum annealers, with D-Wave being the most prominent of them.
D-Wave uses the QUBO formulation to implement the final Hamiltonian~\cite{venegas2018cross}, defined by the following objective function: 
\begin{equation}
E(\bm{x}) = \bm{x}^\T Q\bm{x},
\end{equation}
where $\bm{x}=\begin{bmatrix}x_1 & \hdots & x_N\end{bmatrix}^\T\in\{0,1\}^N$ is a binary vector and $Q$ a real-valued $n\times n$-matrix.
Equivalently, we can write
\begin{equation}
E(\bm{x})= \sum_{i=1}^{N}Q_{ii}x_i+ \sum_{i=1}^N\sum_{j=i+1}^N Q_{ij}x_ix_j.
\end{equation}
From this $Q$, we can construct the Hamiltonian suitable for the D-Wave quantum annealing hardware. 
The code used to implement the code is available online~\cite{TNO_Quantum_N1_problem_2023}.

\subsection{QUBO formulation for the \textit{N-1} problem}
We already saw that the \textit{N-1} problem is easily solved classically for $k=1$. 
We only consider the $k>2$ case in the remainder of this section. 
The formulated QUBO must decide if a specific edge $e$ is \textit{N-1} compliant.
This section gives a high level overview of the QUBO, while \autoref{app:QUBO_formulation} gives the details. 

Interpreting the \textit{N-1} problem as an optimization problem, we have to search for a configuration of edges that minimizes the number of switches, while adhering to the load-flow constraints.
What remains is encoding the graph structure and the switches in binary variables and formulate an objective function $H_\text{obj}$, which is minimized when the number of switches is minimized.
We additionally need penalty terms $P_\text{tree}$ and $P_\text{lf}$ to filter out graphs that are not spanning trees or that do not comply with the load-flow constraints, respectively.

The penalty term $P_\text{tree}$ filters non-spanning tree configurations.
To accomplish this, the penalty term uses the fact that every tree can be represented as a rooted tree.
Rooted trees have properties that translated to the QUBO formalism, which helps constructing a penalty term $P_\text{tree}$.

The penalty term $P_\text{lf}$ filters out spanning trees that do not comply with load-flow constraints. 
We can check the load-flow constraints by solving a (complex) linear system $A\bm{x}=\bm{b}$, followed by checking if $\bm{x}$ is within some bounds.
The load-flow penalty term encodes $\bm{x}$ such that it is always within the bounds and then computes the squared residual $\|A\bm{x}-\bm{b}\|^2_2$.
As a result, the load-flow penalty term vanishes precisely if $x$ is within the bounds and satisfies the load-flow constraints. 
In all other cases $\|A\bm{x}-\bm{b}\|^2_2$ is positive.

As the load-flow penalty function $P_\text{lf}$ works on a spanning trees. 
$P_\text{lf}$ therefore needs information on the specific spanning tree from $P_\text{tree}$.
This information can be transferred by coupling the variables from $P_\text{tree}$ to those of $P_\text{lf}$.
Coupling these variables comes at the cost of extra auxiliary variables for which we need a third penalty term $P_\text{aux}$. 

\autoref{sec:QUBO:spanning_tree} gives details on the $P_\text{tree}$ term, whereas \autoref{sec:QUBO:load_flow} gives details on the $P_\text{lf}$ term. 
\autoref{eq:QUBO:penalty_aux} describes the $P_\text{aux}$ term. 
Putting everything together results in a QUBO problem with the following structure:
\begin{equation}
\min H_\text{obj} + P_\text{tree} + P_\text{lf} + P_\text{aux}.
\end{equation}

\subsection{Method}
To test the performance of the QUBO formulation and understand its behavior, the QUBO formulation was used to solve two small \textit{N-1} problems with both simulated and quantum annealing.
\autoref{fig:qa_larger_solution_graph} (left) and \autoref{fig:qa_smaller_graph} show the input graphs used to test the quantum annealing algorithm for solving the \textit{N-1} problem.
For both graphs, the mathematical QUBO formulation was implemented and solved using D-Wave's dimod, dwave.systems and dwave.samples packages. 

For the quantum annealing approach, the QUBO was sampled with the D-Wave Advantage system.
We used 500 reads with a \unit[1000]{$\mu$s} annealing time and kept all other settings at default values.
This resulted in a total QPU time of \unit[0.565]{s}.
The penalty factors in $P_\text{tree}$ were chosen by checking whether an optimum could be found or not manually (using a brute force approach).
The results obtained by QA were further post-processed by using a steepest descent method implemented by D-Wave.
This post-processing improves the results by finding the nearest local optimum. 

Simulated annealing was performed using the D-Wave implementation.
For each optimization procedure 100 reads were done with 10,000 sweeps and 20 sweeps per beta.
Again, the penalty factors were chosen by hand. 

\subsection{Results and Discussion}
\begin{figure}[tbp]
\centering
\includegraphics[width=\linewidth]{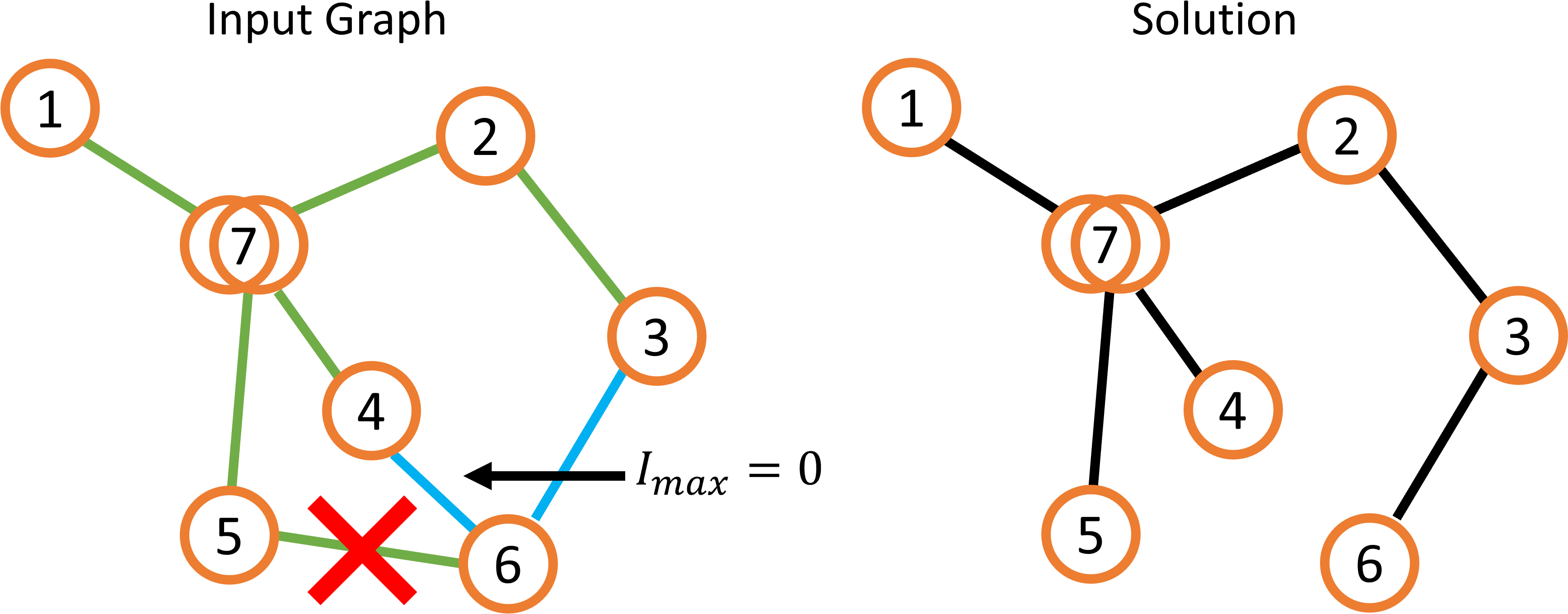}
\caption{Left: Input graph of the small \textit{N-1} problem. 
Single orange circles represent MSR nodes, while the double circle represents the OS node. 
The green edges represent active edges, while the blue edges represent inactive edges. 
The red cross shows which the failing edge. 
Right: The optimal solution to this \textit{N-1} problem, with the $(3,6)$ edge turned on.}
\label{fig:qa_larger_solution_graph}
\end{figure}

First the graph shown in \autoref{fig:qa_larger_solution_graph} was considered.
\autoref{sec:data_small_experiment} shows the exact numeric values  used for the load-flow computations for this graph.
The $I_\text{max}$ value of edge $(4, 6)$ was set to zero, such that there is exactly one optimal solution.

Using simulated annealing as a backend to the algorithm, the optimal solution, shown in \autoref{fig:qa_larger_solution_graph} right, could be found.
To do so, values for the hyper-parameters of the penalty terms were hand picked.
This results shows that the QUBO formulation can indeed find a solution to the \textit{N-1} problem.

Running this same QUBO problem with quantum annealing hardware did not result in a solution for the \textit{N-1} problem.
The main problem was that the quantum annealer did not differentiate between load-flow compliant and non-load-flow compliant solutions.
Possible explanation is that the penalty terms were set suboptimal, something a quantum annealer is quite sensitive to. 
A more riguous choice of these penalty terms might therefore enable QA to find a solution. 
A second possible explanation is that the D-Wave machines have insufficient resolution to accurately reproduce the complete QUBO formulation, which in turn might relate to the sensitivity of the penalty terms in QA. 
Finally, the physically feasible annealing time might be too short to obtain the optimal solution. 
Theoretically, the quality of the solution should increase with longer annealing times.
With current noisy hardware however, we did observe a significant decrease in solution quality for longer annealing times, as qubits decohere. 

\begin{figure}[tbp]
\centering
\includegraphics[width=0.3\linewidth]{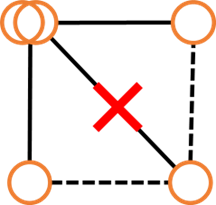}
\caption{Small \textit{N-1} graph.}
\label{fig:qa_smaller_graph}
\end{figure}

To investigate how the penalty terms of the QUBO influence the performance of the QA method, we considered a smaller network, shown in \autoref{fig:qa_smaller_graph}, and omitted the load-flow constraints (the $P_\text{lf}$ and $P_\text{aux}$ penalty terms).
The smaller problem size made it easier to judge the performance of the QUBO.
Note that the penalty term $P_\text{tree}$ consists of multiple terms to cover all possible trees. 
\autoref{app:QUBO_formulation} has more details on this penalty term. 

\begin{figure}[tbp]
\centering
\includegraphics[width=\linewidth]{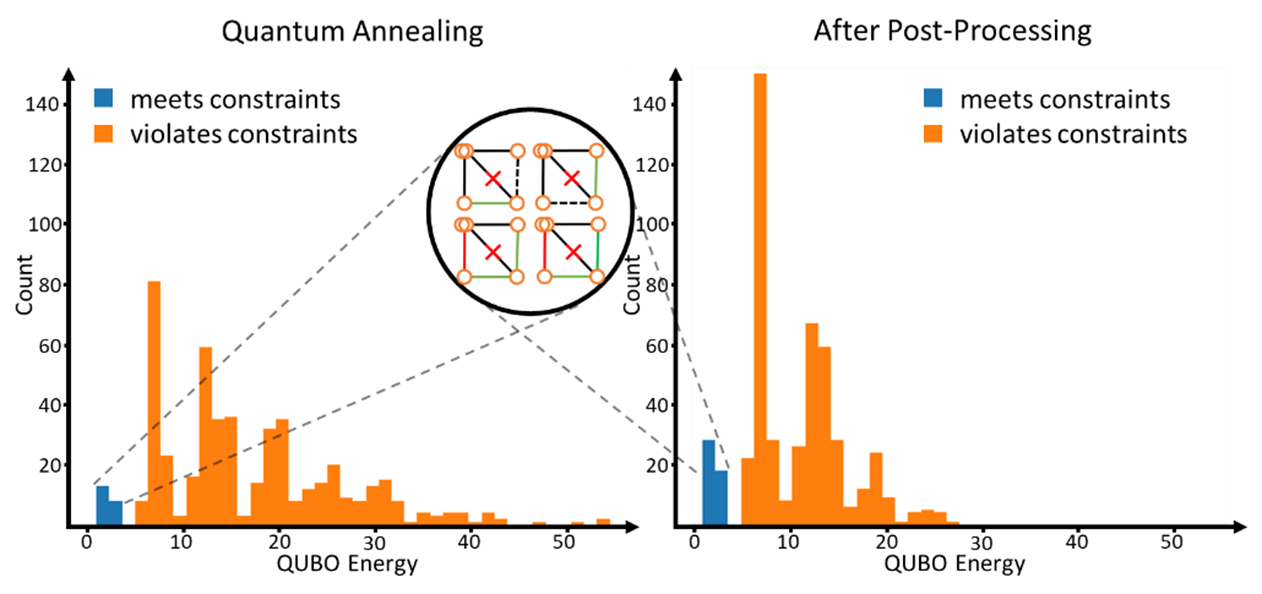}
\caption{Results of QA before (left) and after post-processing with steepest descent (right). The horizontal axis shows the QUBO energy (lower is better), which is equal to the number of switches when the constraints are met. 
The vertical axis shows the number of times a certain energy was found (out of a total of $500$). 
The blue bars indicate runs that meet all constraints (and are guaranteed to be spanning trees), while the orange bars violate at least one constraint.}
\label{fig:qa_histograms}
\end{figure}

In a trial-and-error approach, we found that optimizing the penalty terms one by one, eventually resulted in the optimal solution when running the algorithm on quantum annealing hardware.
\autoref{fig:qa_histograms} (left) shows these results. 
Note that only $13$ out of the $500$ runs gave the optimal solution, and in total only $21$ runs found a spanning tree. 
The QUBO can thus find spanning trees, even on actual hardware. 
Better hardware, combined with tuning the penalty terms, can improve the precision even further, even when considering larger problem instances. 

Apart from better, hardware, classical post-processing can also help to find better solutions. 
A steepest descent algorithm on the measurement results assures that the solutions are in a local minima. 
\autoref{fig:qa_histograms} (right) shows these results after post-processing. 
Interestingly, the solution is now find $28$ times, and a total of $46$ runs gave a spanning tree. 
Post-processing improved the results by more than a factor $2$. 
The post-processing step is efficient and hence, the computational complexity of the algorithm does not increase, while improve the probability of finding the optimal solution significantly.

In conclusion, we have showed that the \textit{N-1} QUBO formulation can find optimal reconfigurations using simulated annealing.
Furthermore, we showed that for small graphs, quantum annealing hardware can help find spanning tree configurations with a minimal amount of switches and that the probability of finding the optimal solution can be improved with a simple steepest-descent post-processing step. 

\section{Photonic quantum computing approach}\label{sec:photonics}
Photonic quantum computers, as the name suggests, make use of photons to perform computations. 
A photon is an elementary particle that is a quantum of the electromagnetic field. 
In common terms, a photon is a light particle. 
Photonic quantum computers have photons as input and let them interact via beamsplitters, phase shifters and loops to eventually detect the photons as the quantum measurement.

\subsection{Universal photonic quantum computers}
Often, photonic quantum computers work with quantum states that are superpositions of, for instance, the number of photons in a state, or superpositions of the input channels a single photon originates from. 
Examples of photon states include Gaussian states and GKP states~\cite{Pennylane2022}.
Current photonic quantum computers do not yet admit universal operations, though approaches exist to make them universal.

As photons seldom interact with each other or their environment, photons, can remain stable for a long time.
The disadvantage, however, is that performing operations on pairs of photons becomes harder.
Moreover, the longer a photon has to travel through a quantum computer, the higher the chances are that a photon gets lost (i.e. absorbed by the environment).
Photon losses limit the maximum length of photon paths, and consequently the amount of operations that can be performed on them in photonic quantum computing architectures.
Finally, it is hard to, repeatedly and fast-paced, produce the right photonic input states~\cite{Takeda2019, Slussarenko2019}.

For the \textit{N-1} problem we therefore consider a special instance of photonic quantum computers that has already seen implementations: the Gaussian boson sampler. 

\subsection{Gaussian boson samplers for the \textit{N-1} problem}
Gaussian Boson Sampling (GBS) is a special-purpose photonic platform that can be used to perform sampling tasks that can not be performed on classical computers~\cite{Hamilton2017, Kruse2019}. 
Several GBS algorithms have been developed over the last couple of years.
The best known are graph optimization, molecular docking, graph similarity, point processes and quantum chemistry~\cite{Bromley2020}.

Since these are closely related to the \textit{N-1} problem, we have studied them in more detail in this work.

All of these algorithms could only solve a small part of the \textit{N-1} problem at most, and therefore were not being suitable.
Apart from the above algorithms, we also developed a more customized algorithm.
The algorithm could efficiently check that given a graph with two -no more, no less- connect parts, whether the addition of an extra edge between two nodes in the graph would make the graph one that is fully connected.
This check is relevant when looking for a switch that activates an inactive edge in case of an edge failure. 
The two graph parts are embedded into two disjunct parts of the device. Running the algorithm consists of sending photons from one part of the graph and measuring the state at the other parts of the device.
Photons will be found here only if an edge exist between the two parts.

The potential of this algorithm for the \textit{N-1} problem is limited however, as we found no way to implement the load-flow computations on Gaussian boson samplers. Moreover, to solve the \textit{N-1} problem, we wish to probe multiple reconfigurations in superposition. This particular device and other NISQ photonic devices seem unsuitable for this task. They are usually used for extracting properties of a single graph instead of checking properties of multiple graphs in parallel.

% \subsection{Future prospects of photonic quantum computing for \textit{N-1}}
% This leads to the conclusion that for photonic quantum computers to be relevant for the \textit{N-1} problem, breakthroughs have to be made to realize universal quantum computers.
% As an universal architecture that might be of special interest for graph problems in the future, we would like to point out the measurement-based quantum computing architecture.

% Measurement-based quantum computing continuously measures photons from one highly entangled state. The measurement results are then used to slightly tune the measurements in the next iteration. In this way the computation is done by changes the measurement basis instead of changing the photon states with specified photon operations. This paradigm, therefore, is very different from the gate-based paradigm. The promise is that if the photons can be produced fast enough and the subsequent measurements can be changed timely, this would allows for a relatively small device.

% The highly entangled quantum state used in measurement-based photonic quantum computing is closely related to graph theory. Therefore, we expect that, graph problems will be particularly interesting to study with the help of these devices. Following developments towards universal photonic measurement-based quantum computing is therefore advisable.

\section{Conclusion}\label{sec:conclusion}
In this work, we considered three different quantum approaches to solve the \textit{N-1} problem encountered in power grid robustness. 
\textit{N-1} security ensures that valid reconfigurations exist in power grids in case of single-link failures. 
Classical algorithms can find such reconfigurations for many simple cases, but struggle with the few harder cases. 
As slight improvements in the complexity of the problems can have great impact in practice, we explore the potential of gate-based quantum computers, quantum annealing devices, and photonic quantum computers. 

For the gate-based quantum computing approach, we used a variant of Grover's search algorithm.
We made a grey-box implementation to mark the valid reconfigurations and counted only the number of calls to the load-flow computation. 
We found a quadratic improvement in the number of calls to load-flow computations with respect to the classical algorithm. 
Due to the large circuit depth of the algorithm, the required number of qubits and the required number of 2-qubit gates, it seems unlikely that this implementation will achieve quantum advantage in the coming five years, unless significant advances in the hardware technology happen.
However, in implementing the algorithm, techniques have been developed that will be useful in future graph-related quantum computing implementations.  

For the quantum annealing approach, we formulated a QUBO formulation for the problem and used that to find valid reconfigurations. 
The QUBO includes a term that minimizes the number of switchovers, hence a single algorithm can directly work for all allowed number of switchovers. 
The QUBO also includes additional terms to assure that the output is a spanning tree and that the load-flow constraints are satisfied. 
We tested the QUBO formulation with simulated annealing and found that the algorithm does indeed find a valid reconfiguration.
We next solved a simpler version of the problem on quantum annealing hardware and showed that current hardware already can help in this~\textit{N-1} problem. 

Different attempts were made to leverage the non-universal photonic quantum computer. 
The main explored route was to use boson scattering effects to quickly check graph connectivity.
Due to the difficulty of performing this operation in superposition, and the fact that the load-flow check also checks graph connectivity, an implementation for this approach was not made. 
 
In the current work, we assumed a classical algorithm for the load-flow computations, yet, more efficient quantum versions might exist. 
These algorithms can then further help solving the \textit{N-1} problem faster.
In settings where the decision version of the \textit{N-1} problem provides enough information, variational algorithms may be of more interest. 
With these variational algorithms, we optimize certain parameters of a quantum circuit. 
Attention may also be shifted away from the \textit{N-1} problem, looking instead for other problems relevant to the Distribution System Operator community and suitable for quantum computing, such as: fast large-scale reconfigurations for important substation/transformer outages, simulating thermodynamic properties of electricity cables, resource scheduling, or power flow analysis. 

\printbibliography

\appendix

\section{Explicit QUBO formulation}\label{app:QUBO_formulation}
In this section we explicitly derive and present the QUBO used to solve the \textit{N-1} problem with a quantum annealer.
We will do so in three steps:
First, we construct a QUBO that searches for spanning trees with minimal switches.
Second, we formulate a QUBO that checks for load flow constraints for a given graph.
Third, we combine the two QUBOs to obtain the \textit{N-1} QUBO formulation.

\subsection{General QUBO formulation tools}
\label{sec:qubo_tools}
\subsubsection{Penalty functions}
QUBO formulations are unconstrained, yet many interesting problems require constraints.
Including a penalty function in the objective overcomes this restraint. 
A pseudo-Boolean function $P:\{0,1\}^n\to\R$ is called a penalty function of a constraint if and only if it is non-negative and is zero if and only if the constraint is met.
By adding such a penalty function to the objective, solutions that violate a constraint become unfavorable.

\noindent\textbf{Linear equality constraints}\\
For constants $a_i,b\in\R$ and binary variables $x_i\in\{0,1\}$, a linear equality constraint has the form:
\begin{equation}
\sum_i a_i x_i = b.
\end{equation}
Linear equality constraints have the following standard penalty function:
\begin{equation}
P(\bm{x})= \left(\sum_i a_i x_i - b\right)^2.
\end{equation}

\noindent\textbf{Pairwise degree reduction}\\
If problem formulations contain terms of higher degree than $2$, auxilliary binary variables can help reduce the overall degree. 
For pairwise degree reduction, a quadratic $x_i x_j$ term is substituted by a new variable $z$ and the following penalty function is added to the objective:
\begin{equation}
P(x_i,x_j,z) = x_i x_j-2z(x_i+x_j )+3z.
\end{equation}
Note that $P(x_i,x_j,z)$ is non-negative and $P(x_i,x_j,z)=0$ if and only if $x_i x_j=z$.

\subsubsection{Discrete variables}
Let denote $\delta_x^i$ the indicator variable for $i\in I$, i.e.,
\[
\delta_x^i = \left\{
\begin{array}{ll}
1, & \text{if } x=i,\\
0, & \text{if } x\not=i.\\
\end{array}\right.
\]
Such a discrete variable requires an encoding enforced by a penalty term $P(\delta_x)$.

\noindent\textbf{One-hot encoding}\\
The most simple encoding of a discrete variable is the one-hot encoding.
In this encoding, a binary variable $x_i\in\{0,1\}$ is introduced for each discrete option $i\in I$.
Therefore, $x_i=1$ for exactly one $i$.
Adding the following penalty term to the objective function enforces this behavior
\begin{equation}
P_\text{oh}=\left(1-\sum_ix_i \right)^2.
\end{equation}
The penalty term above is non-negative and zero if and only if exactly one $x_i$ is 1.
After adding this penalty term, one simply substitutes $\delta_x^i=x_i$.

\noindent\textbf{Domain-wall encoding}\\
A more efficient encoding for discrete variables is the domain-wall encoding~\cite{chancellor2019domain}.
For this encoding $|I|-1$ variables are used, labeled $x_1,x_2,…x_{(|I|-1)}$. Furthermore, for notation purposes we use $x_0=0$ and $x_{|I|}=1$.
The bitstring $x_0\ x_1\ \cdots \ x_{|I|}$ is constrained by allowing for precisely one place where two adjacents bits have different values.
Hence, both $00111$ and $00001$ are valid bitstring, whereas $01011$ is not.
This behavior is encoded using the following penalty term:
\begin{equation}
P_\text{dw} = \sum_{i=1}^{|I|-2}x_i(1-x_{i+1}).
\end{equation}
This term counts the number of 1-0 substrings, which is zero for all valid bitstrings.
The penalty term is zero if and only if the domain-wall constraint is met and positive otherwise. 
After adding this penalty term, one replaces the discrete indicator variables with the locations of the domain wall: $\delta_x^i=x_i-x_{i-1}$.

\subsection{QUBO for spanning trees with minimal switches}
\label{sec:QUBO:spanning_tree}
The goal of this section is to write a QUBO formulation that finds the spanning trees with a minimal amount of switches.
 
\noindent\textbf{Problem}\\
Given a graph $G=(V,E=E_a\cup E_i)$, with a set of active edges $E_a$ and inactive edges $E_i$, find a spanning tree $T=(V,E_\text{tree})$ of $G$, such that the symmetric difference $k=|E_\text{tree}\, \Delta \,E_a |$ is minimized.

\begin{figure}[tbp]
\centering
\includegraphics[width=0.3\linewidth]{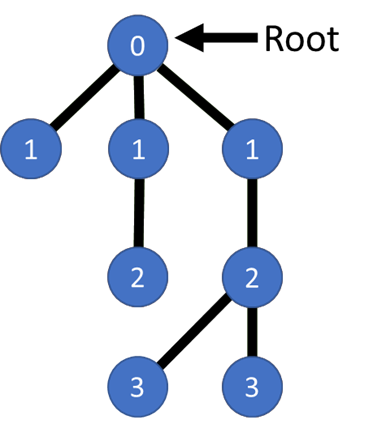}
\caption{A rooted tree. The numbers in the nodes represent the depth of each vertex.}
\label{fig:rooted_tree}
\end{figure}

To translate this problem to the QUBO formulation, we start by constructing a QUBO problem that is minimized if and only if the solution represents a spanning tree.
Interestingly, rooted trees have an easier QUBO formulation than trees in general. 
In such trees, one single vertex is designated as the root. 
Each vertex also has a depth equal to the length of the longest path starting from the root to that vertex. It turns out that finding a specifically structured spanning tree is more easily translated to a QUBO formulation than a general tree.
Given a tree, we obtain a rooted tree by simply assigning one of the vertices as root, vice versa we can omit the root label to obtain a normal tree from a rooted one~\cite{bender2010lists}. 
\autoref{fig:rooted_tree} gives an example of a rooted tree. 

A QUBO follows from the rooted tree structure by constructing penalty functions for the following constraints (see also~\cite{lucas2014ising}):
\begin{itemize}
    \item Each vertex has precisely one depth.
    \item Exactly one vertex is the root vertex.
    \item Every non-root vertex connects to exactly one vertex with a lower depth.
    \item Every vertex has no connection with vertices of the same depth.
\end{itemize}

\subsubsection{Quadratic Unconstrained Discrete Optimization formulation}
We start with reformulating the problem as a quadratic unconstrained discrete optimization problem, and then translate that into a QUBO formulation.

Let $I\le |V|$ be the maximum height of the rooted tree.
For each node $v\in V$, the discrete indicator variable $\delta_{x_v}^i$ represents the depth of the node, with a total of $I$ possible values.
For each edge $\{v,u\}\in E$ a discrete indicator variable $\delta_{y_{vu}}^i$ is introduced, with $2I-1$ options.
If $\delta_{y_{vu}}^i=1$ for $i\in\{0,1,\ldots,I-2\}$ , then edge $\{v,u\}$ is in layer $i$ and $u$ is closer to the root then $v$.
If $\delta_{y_{vu}}^i=1$ for $i\in\{I-1,I,\ldots,2I-2\}$, then edge $\{v,u\}$ is in layer $i$ and $v$ is closer to the root then $u$.
If $\delta_{y_{vu}}^{2I-1}=1$, then edge $\{v,u\}$ is not part of the tree.

Note, that with these discrete variables, every vertex can have exactly one depth. Thus, if we find and assignment of variables with exactly one root node (depth=0) and we make sure that each non-root node connects with exactly one node in the layer above and that there are no connections between vertices with the same depth, then we have found a (rooted) tree.

\textbf{Penalties}

\noindent\textbf{One root}
\begin{equation}
P_\text{root}=(1-\sum_{v\in V}\delta_{x_v}^0 )^2.
\end{equation}
\textbf{Connectivity}\\
Every non-root vertex connects to exactly one vertex with a lower depth.
\begin{equation}
P_\text{con}
=
\sum_{v\in V}\sum_{i=1}^{I-1}
\left(
\delta_{x_v}^i
-\sum_{u\mid\{v,u\}\in E}\delta_{y_{vu}}^{i-1}
- \sum_{u\mid\{u,v\}\in E} \delta_{y_{uv}}^{I+i-2}
\right)^2 .
\end{equation}
\textbf{Behavior of indicators}\\
If an edge has depth $i$, then the incident nodes should be in layer $i$ and $i+1$.
\begin{equation}
P_\text{ind}
=
\sum_{\{v,u\}\in E}\sum_{i=0}^{I-2}
\delta_{y_{vu}}^i \left(2-\delta_{x_v}^{i+1}-\delta_{x_u}^i \right)
+ \delta_{y_{vu}}^{I-1+i}\left(2-\delta_{x_v}^i-\delta_{x_u}^{i+1}\right).
\end{equation}
Note that the penalty above also implies that every vertex has no connection to a vertex with the same depth.

\textbf{Objective}
The objective is to minimize the number of needed switches.
\begin{equation}
H_\text{obj}
=
\sum_{\{v,u\}\in E_a}\delta_{y_{vu}}^{2I-1}
+\sum_{\{v,u\}\in E_i}1-\delta_{y_{vu}}^{2I-1}.
\end{equation}
The whole problem can be formulated as the following unconstrained discrete optimization problem
\begin{equation}
\min H_\text{obj} + P_\text{root} + P_\text{con} + P_\text{ind}.
\end{equation}

\subsubsection{QUBO Formulation}
Using one of the encodings introduced in Section~\ref{sec:qubo_tools}, binary variables can act like discrete variables. 
We provide an example using the domain-wall encoding.
\begin{align}
P_\text{dw}&=
\sum_{v\in V}\sum_{i=0}^{I-3}x_{v,i} (1-x_{v,i+1})
+\sum_{\{v,u\}\in E}\sum_{i=0}^{2I-3}y_{vu,i}(1-y_{vu,i+1}),\\
P_\text{root}&=
\left(1-\sum_{v\in V}x_{v,0}\right)^2,\\
P_\text{con}&=
\sum_{v\in V}\sum_{i=1}^{I-1}
\left(x_{v,i}-x_{v,i-1}
-\!\!\!\!\!\!\!\sum_{u:\{v,u\}\in E}y_{vu,i-1}-y_{vu,i-2}
-\!\!\!\!\!\!\!\sum_{u:\{u,v\}\in E}y_{uv,I+i-2}-y_{uv,I+i-3}\right)^2,\\
P_\text{ind}&=
\sum_{\{v,u\}\in E}\sum_{i=0}^{I-2}
(y_{vu,i}-y_{vu,i-1})(2-x_{v,i+1}+x_{v,i}-x_{u,i}+x_{u,i-1} )\\
&\qquad+(y_{vu,I+i-1}-y_{vu,I+i-2})(2-x_{v,i}+x_{v,i-1}-x_{u,i+1}+x_{u,i} ),\\
H_\text{obj}&=
\sum_{\{v,u\}\in E_a}1-y_{vu,2I-2} + \sum_{\{v,u\}\in E_i}y_{vu,2I-2}.
\end{align}
The complete QUBO problem is then given by
\begin{equation}
\min H_\text{obj}+ P_\text{tree},
\end{equation}
where
\begin{equation}
P_\text{tree}=
\lambda_\text{dw} P_\text{dw}
+\lambda_\text{root}P_\text{root}
+\lambda_\text{con}P_\text{con}
+\lambda_\text{ind}P_\text{ind}.
\end{equation}
In the equation above $\lambda_\text{dw}$,  $\lambda_\text{root}$, $\lambda_\text{con}$ and $\lambda_\text{ind}$ are real valued positive constants called penalty factors (or hyper parameters).
They determine the relative importance of each penalty term.

\subsection{QUBO formulation for penalizing non-load-flow compliant trees}
\label{sec:QUBO:load_flow}
In this section, we will construct a QUBO formulation that penalizes a non-load-flow compliant tree.
We first give a mathematical formulation of the problem (see also~\cite{fritschy2018checking}) and then approximate this continuous problem be a discrete one.

Let $G=(V,E)$ be a graph with vertex set $V=V_\text{OS}\cup V_\text{MSR}$ and edge set $E$.
Each node $v\in V_\text{MSR}$ has the following attributes:
\begin{itemize}
    \item A real positive value $U_n^\text{min}>0$,
    \item A real positive value $U_n^\text{max}\ge U_n^\text{min}$,
    \item A real positive value $R_n$.
\end{itemize}

Each node $n \in V_\text{OS}$ has a complex valued attribute $U_n$.
Each edge $\{n,m\}\in E$ has he following attributes:
\begin{itemize}
    \item A real positive value $I_{\{n,m\}}^\text{max}>0$,
    \item A complex value $Z_{\{n,m\}} \in \C \setminus\{0\}$.
\end{itemize}

We say that $G$ complies with the load-flow equations if for all MSR nodes $v\in V_\text{MSR}$ there exists at least one combination of $U_n\in\C$ such that the following equations are met:
\begin{eqnarray}
\frac{U_n}{R_n} &= \sum_{m\in N(n)}\frac{U_m-U_n}{Z_{\{n,m\}}}, &\forall n\in V_\text{NMR},\\
&U_n^\text{min}\le |U_n| \le U_n^\text{max}, &\forall n\in V_\text{NMR},\\
&\left|\frac{U_m-U_n}{Z_{\{n,m\}}} \right| \le I_{\{n,m\}}^\text{max}, & \forall \{n,m\}\in E,
\end{eqnarray}
where $N(n)$ is the set of neighbors of $n$.
If no such solution exists, then $G$ does not comply with the load-flow equations.

\textbf{Splitting and simplifying the load-flow equations}\\
The imaginary voltages and currents are at least an order of magnitude smaller compared to their real counterparts.
Furthermore, it is assumed that the voltages are positive.
Therefore, we can use the following approximation $|U_n |\approx \Re(U_n)$, to produce
\begin{eqnarray}
\frac{U_n}{R_n} = \sum_{m\in N(n)}\frac{U_m-U_n}{Z_{\{n,m\}}} &\forall n\in V_\text{NMR},\\
U_n^\text{min} \le \Re(U_n) \le U_n^\text{max}&\forall n\in V_\text{NMR},\\
-I_{\{n,m\}}^\text{max} \le \Re\left(\frac{U_m-U_n}{Z_{\{n,m\}}} \right) \le I_{\{n,m\}}^\text{max}&\forall \{n,m\} \in E.
\end{eqnarray}

Furthermore, we can split the first equation into a real part and an imaginary part, which gives:
\begin{eqnarray}
\frac{\Re(U)_n}{R_n}
= \sum_{m\in N(n)} \Re(U_m-U_n)\Re\left(\frac{1}{Z_{\{n,m\}}}\right) - \Im(U_m-U_n)\Im\left(\frac{1}{Z_{\{n,m\}}} \right)
&\forall n \in V_\text{NMR},\\
\frac{\Im(U_n)}{R_n}
=\sum_{m\in N(n)} \Re(U_m-U_n )\Im(\frac{1}{Z_{\{n,m\}}})+\Im(U_m-U_n )\Re\left(\frac{1}{Z_{\{n,m\}}}\right)
&\forall n \in V_\text{NMR},\\
U_n^\text{min} \le \Re(U_n) \le U_n^\text{max}&\forall n\in V_\text{NMR},\\
-I_{\{n,m\}}^\text{max} \le \Re\left(\frac{U_m-U_n}{Z_{\{n,m\}}} \right) \le I_{\{n,m\}}^\text{max}&\forall \{n,m\} \in E.
\end{eqnarray}

\subsubsection{Discrete approximation}
The system of equations above contains real-valued variables $\Re(U_n)$ and $\Im(U_m)$ for each $n \in V_\text{MSR}$.
These are incompatible with the QUBO formulation and therefore with quantum annealing.
To overcome this problem each variable will be discretized.

Next, we choose $K,L,J \in N$ and introduce the following pseudo-Boolean functions:
\begin{table}[htbp]
\centering
\begin{tabular}{lll}
\toprule
Name	   &Domain	    &Codomain\\
\toprule
$U_n^\R$   &$\{0,1\}^K$	&$[U_n^\text{min},U_n^\text{max} ]$\\
$U_n^\I$   &$\{0,1\}^L$	&$[-0.1U_n^\text{min},0.1U_n^\text{min} ]$\\
$I_{\{n,m\}}$	&$\{0,1\}^J$	&$[-I_{\{n,m\}}^\text{max},I_{\{n,m}\}^\text{max} ]$\\
\bottomrule
\end{tabular}
\end{table}

Defined by
\begin{eqnarray}
U_n^\R (\bm{u}_n^\R )=U_n^\text{min}+(U_n^\text{max}-U_n^\text{min})2^{-K}\left(1+\sum_{k=1}^K2^k u_{n,k}^\R \right),\\
U_n^\I (\bm{u}_n^\I)=-0.1U_n^\text{min}+0.1U_n^\text{min}2^{-L} \sum_{l=1}^L 2^l u_{n,l}^\I,\\
I_{\{n,m\}} (\bm{i}_{\{n,m\}} )=-I_{\{n,m\}}^\text{max}+-I_{\{n,m\}}^\text{max}2^{-J}  \left(1+\sum_{j=1}^J2^j i_{\{n,m\},j}\right).
\end{eqnarray}

For notation purposes, we introduce $\alpha_n=R_n^{-1}$, $\beta_{\{n,m\}}=\Re\left(Z_{\{n,m\}}^{-1}\right)$ and $\gamma_{\{n,m\}}=\Im\left(Z_{\{n,m\}}^{-1}\right)$.
Note that these are all real-valued constants.
With these constants and the above pseudo-Boolean functions, we can approximate the load-flow equations with the following equations:
\begin{eqnarray}
\alpha_n U_n^\R (\bm{u}_n^\R)=\sum_{m\in N(n)}\beta_{\{n,m\}}(U_m^\R (\bm{u}_m^\R )-U_n^\R (\bm{u}_n^\R ))- \gamma_{\{n,m\}}(U_m^\I (\bm{u}_m^\I )-U_n^\I (\bm{u}_n^\I ))    &\forall n \in V_\text{NMR},\\
\alpha_n U_n^\I (\bm{u}_n^\I)=\sum_{m\in N(n)}\gamma_{\{n,m\}} (U_m^\R (\bm{u}_m^\R )-U_n^\R (\bm{u}_n^\R ))+ \beta_{\{n,m\}} (U_m^\I (\bm{u}_m^\I )-U_n^\I (\bm{u}_n^\I )) & \forall n \in V_\text{NMR},\\
I_{\{n,m\}} (\bm{i}_{\{n,m\}})=\beta_{\{n,m\}} (U_m^\R (\bm{u}_m^\R )-U_n^\R (\bm{u}_n^\R ))- \gamma_{\{n,m\}} (U_m^\I (\bm{u}_m^\I )-U_n^\I (\bm{u}_n^\I )) & \forall \{n,m\} \in E.
\end{eqnarray}

Note that for some edges, the last constraint can never be violated, as if
\begin{equation}
\max \{U_n^\text{max}-U_m^\text{min},U_m^\text{max}-U_n^\text{min} \}|\beta_{\{n,m\}}|
+0.1(U_n^\text{min}+U_m^\text{min}) |\gamma_{\{n,m\}}|
\le
I_{\{n,m\}}^\text{max},
\end{equation}
then the current can never exceed the limit as long as the voltages are within the bounds.
Since in this formulation the voltages cannot be out of bounds, we can drop the constraints when the equation above holds.
The set $E_p$ contains all problem edges for which this is not the case.
So the last constrained becomes
\begin{equation}
I_{\{n,m\}} (\bm{i}_{\{n,m\}})
=\beta_{\{n,m\}} (U_m^\R (\bm{u}_m^\R )-U_n^\R (\bm{u}_n^\R ))- \gamma_{\{n,m\}} (U_m^\I (\bm{u}_m^\I )-U_n^\I (\bm{u}_n^\I )) \qquad \forall \{n,m\} \in E_p.
\end{equation}

\subsubsection{QUBO formulation}
To construct a QUBO formulation for this problem, we will construct the following penalty functions
\begin{eqnarray}
P_{U^\R}=\sum_{n\in V_\text{NMR}}
\left(\alpha_n U_n^\R (
\bm{u}_n^\R )+\sum_{m\in N(n)}\beta_{\{n,m\}}(U_n^\R (\bm{u}_n^\R )-U_m^\R (\bm{u}_m^\R ))- \gamma_{\{n,m\}} (U_n^\I (\bm{u}_n^\I )-U_m^\I (\bm{u}_m^\I ))\right)^2 ,\\
P_{U^\I}=\sum_{n \in V_\text{NMR}}
\left(\alpha_n U_n^\I (\bm{u}_n^\I )
+\sum_{m\in N(n)}\gamma_{\{n,m\}}(U_n^\R (\bm{u}_n^\R )-U_m^\R (\bm{u}_m^\R ))
+\beta_{\{n,m\}}(U_n^\I (\bm{u}_n^\I )-U_m^\I (\bm{u}_m^\I ))
\right)^2 ,\\
P_I=\sum_{\{n,m\}\in E_p}
\left(I_{\{n,m\}} (\bm{i}_{\{n,m\}})
+\beta_{\{n,m\}}(U_n^\R (\bm{u}_n^\R )-U_m^\R (\bm{u}_m^\R ))
- \gamma_{\{n,m\}} (U_n^\I (\bm{u}_n^\I )-U_m^\I (\bm{u}_m^\I ))
\right)^2.
\end{eqnarray}
The QUBO formulation of the problem is then given by:
\begin{equation}
\min_{\bm{u}^\R,\bm{u}^\I,\bm{i}}
\lambda_{U^\R} P_{U^\R}
+\lambda_{U^\I}P_{U^\I}
+\lambda_I P_I,
\end{equation}
where $\lambda_{U^\R}$, $\lambda_{U^\I}$ and $\lambda_I$ are real positive constants. We then say that the tree complies with the load-flow equations if the minimum of the QUBO formulation above is close to zero.

\noindent\textbf{Limitations}
\begin{enumerate}
    \item $K$, $L$ and $J$ must be sufficiently large to differentiate between a non-optimal solution and a value corresponding to an non-complaint graph.
    \item $P_{U^\R}$ and $P_{U^\I}$ can be interpreted as minimizing the residual of a linear system $A\bm{x}=\bm{b}$. 
    If $A$ has a large condition number, low values of the QUBO can still have large errors. 
    During this project it was observed that large condition numbers exclusively occurred for disconnected graphs.
\end{enumerate}

\subsubsection{N-1 QUBO formulation}
We produce an \textit{N-1} QUBO formulation by linking the tree structure from the tree QUBO formulation to the load-flow QUBO formulation.
The load-flow QUBO as presented in the previous section governs the equation of the entire graph.
Now, it should only take into account edges that are selected.
Note that if we use the domain-wall encoding and an edge $\{n,m\}\in E$ has a depth (and thus is part of the tree), then $y_{nm,2I-2}=1$, otherwise it is $0$.
Hence, $y_{nm,2I-2}$ is a binary indicator variable that shows if an edge is part of the tree and should be taken into account for the load-flow equations.
Hence, the new load-flow penalty terms become

\begin{multline}
P_{U^\R}=\sum_{n\in V_\text{NMR}}\Big(\alpha_n U_n^\R (\bm{u}_n^\R) +\sum_{m\in N(n)} y_{nm,2I-2}\beta_{\{n,m\}} (U_n^\R (\bm{u}_n^\R )-U_m^\R (\bm{u}_m^\R ))\\
-y_{nm,2I-2}\gamma_{\{n,m\}} (U_n^\I (\bm{u}_n^\I )-U_m^\I (\bm{u}_m^\I ))\Big)^2,
\end{multline}
\begin{multline}
P_{U^\I}=\sum_{n\in V_\text{NMR}}\Big(\alpha_n U_n^\I (\bm{u}_n^\I ) +\sum_{m\in N(n)}y_{nm,2I-2} \gamma_{\{n,m\}}(U_n^\R (\bm{u}_n^\R )-U_m^\R (\bm{u}_m^\R ))\\
+ y_{nm,2I-2}\beta_{\{n,m\}}(U_n^\I (\bm{u}_n^\I )-U_m^\I (\bm{u}_m^\I ))\Big)^2 ,
\end{multline}
\begin{multline}
P_I=\sum_{\{n,m\}\in E_p} \Big(I_{\{n,m\}} (\bm{i}_{\{n,m\}} +y_{nm,2I-2}\beta_{\{n,m\}}(U_n^\R (\bm{u}_n^\R )-U_m^\R (\bm{u}_m^\R ))\\
-y_{nm,2I-2}\gamma_{\{n,m\}}(U_n^\I (\bm{u}_n^\I )-U_m^\I (\bm{u}_m^\I )) \Big)^2 .
\end{multline}
Note that this gives 3rd and 4th order terms, which are not allowed in the QUBO formulation.
Hence, we perform pairwise degree reduction to quadratize the equations.
Let $z_{\{n,m\},n}^\R=y_{nm,2I-2} u_n^\R$ and $z_{\{n,m\},n}^\I=y_{nm,2I-2} u_n^\I$, then the quadratized form is given by
\begin{multline}
P_{U^\R}=\sum_{n\in V_\text{NMR}}\Big(\alpha_n U_n^\R (\bm{u}_n^\R) +\sum_{m\in N(n)} \beta_{\{n,m\}} (U_n^\R (\bm{z}_{\{n,m\},n}^\R )-U_m^\R (\bm{z}_{\{n,m\},m}^\R ))\\
-y_{nm,2I-2}\gamma_{\{n,m\}} (U_n^\I (\bm{z}_{\{n,m\},n}^\I )-U_m^\I (\bm{z}_{\{n,m\},m}^\I ))\Big)^2,
\end{multline}
\begin{multline}
P_{U^\I}=\sum_{n\in V_\text{NMR}}\Big(\alpha_n U_n^\I (\bm{u}_n^\I ) +\sum_{m\in N(n)} \gamma_{\{n,m\}}(U_n^\R (\bm{z}_{\{n,m\},n}^\R )-U_m^\R (\bm{z}_{\{n,m\},m}^\R ))\\
+\beta_{\{n,m\}}(U_n^\I (\bm{z}_{\{n,m\},n}^\I )-U_m^\I (\bm{z}_{\{n,m\},m}^\I ))\Big)^2 ,
\end{multline}
\begin{multline}
P_I=\sum_{\{n,m\}\in E_p} \Big(I_{\{n,m\}} (\bm{i}_{\{n,m\}} +\beta_{\{n,m\}}(U_n^\R (\bm{z}_{\{n,m\},n}^\R )-U_m^\R (\bm{z}_{\{n,m\},m}^\R ))\\
-y_{nm,2I-2}\gamma_{\{n,m\}}(U_n^\I (\bm{z}_{\{n,m\},n}^\I )-U_m^\I (\bm{z}_{\{n,m\},m}^\I )) \Big)^2 .
\end{multline}
Lastly, we add a penalty term for the auxiliary variables $\bm{z}^\R$ and $\bm{z}^\I$.
\begin{multline}
P_\text{aux}=\lambda_\text{aux} \sum_{\{n,m\}\in E} \sum_{k=1}^Ky_{nm,2I-2} u_{nk}^\R-2z_{\{n,m\}nk}^\R (y_{nm,2I-2}+u_{nk}^\R )+3z_{\{n,m\}nk}^\R \\
+\sum_{l=1}^L y_{nm,2I-2} u_{n,l}^\I-2z_{\{n,m\}nl}^I (y_{nm,2I-2}+u_{nl}^\I )+3z_{\{n,m\}nl}^\I.
\label{eq:QUBO:penalty_aux}
\end{multline}
Which produces the final \textit{N-1} QUBO formulation:
\begin{equation}
\min H_\text{obj}+P_\text{tree}+P_\text{lf}+P_\text{aux}.
\end{equation}

\section{Load-flow data for the small dataset}
\label{sec:data_small_experiment}

\begin{table}[htbp]
    \centering
    \begin{tabular}{llllll}
    \toprule
    ID	&Type	&$U^\text{NOM}$	&Load	&$U^\text{min}$	&$U^\text{max}$\\
    \bottomrule
    1	&MSR    &10500	        &992.25i	&9800	&11000\\
    2	&MSR    &10500	        &2721.659i	&9800	&11000\\
    3	&MSR    &10500	        &2721.659i	&9800	&11000\\
    4	&MSR    &10500	        &2746.796i	&9800	&11000\\
    5	&MSR    &10500	        &2735.964i	&9800	&11000\\
    6	&MSR  	&10500	        &793842.2+378992.557i	&9800	&11000\\
    7	&OS	    &10500	        &0	        &10500	&10500\\
    \bottomrule
    \end{tabular}
    \caption{Nodes}
    \label{tab:nodes}
\end{table}

\begin{table}[htbp]
    \centering
    \begin{tabular}{llll}
    \toprule
    Edge	&$Z$	&$I^\text{max}$	&Active\\
    \bottomrule
    $\{1,7\}$ &0.0189+0.0309i	&557.75	&Yes\\
    $\{2,3\}$ &0.060132+0.052799i	&349.2	&Yes\\
    $\{2,7\}$ &0.5i	&999	&Yes\\
    $\{3,6\}$ &(1+i) $10^{-6}$	&999	&No\\
    $\{4,6\}$ &0.059778+0.053095i	&0	&No\\
    $\{4,7\}$ &0.5i	&999	&Yes\\
    $\{5,6\}$ &0.059136+0.0528i	&378.3	&Yes\\
    $\{5,7\}$ &0.5i	&999	&Yes\\
    \bottomrule
    \end{tabular}
    \caption{Edges}
    \label{tab:edges}
\end{table}
\end{document}